\documentclass[12pt]{article}
\usepackage{graphicx}
\usepackage{amssymb}

\title{The Flat Deuteron}
\author{Omar Y\'epez\\
Department of Chemistry, Memorial University of Newfoundland, \\
St. JohnÕs, NF, A1B 3X7
Canada.}
\begin{document}
\maketitle
\begin{abstract}
The new model \cite{yepez} was applied to the femtometer toroidal structures found for the deuteron in \cite{forest}. It was possible to relate the magnetic moment and the energy of the particle to the torus geometric parameters. Excellent agreement between the magnetic moment of the deuteron and the theory developed for a time consuming intersecting 3D torus in a 2D space was found. It was possible to discriminate between which structures present a magnetic moment and which one would produce an anapole moment. This supports the concept that there are orientations of the torus that make it lose its magnetic moment, which is the key to understanding the EPR paradox. The mass energy of the particle changed with each structure, but the mass density was constant, which is consistent with the tension exerted on the space as the source of the particle's "mass" or more appropriately said the particle's density.
\end{abstract}
\section{Introduction}
Due to the lack of a better model or explanation, the assumption that the quantum particle is just a wave, described by the wave equation, has prevailed for decades. Since the electron has shown a point particle behavior \cite{enigmatic49}, it has been assumed that what is waving is a charged point particle in a mysterious potential well. This dancing in its "orbit" (the mysterious potential well) produces an orbital angular momentum and a orbital dipolar magnetic moment, the ratio of which is related to the particle  magnetic moment \cite{eisberg318}. Therefore, the orbital angular momentum is as real as the point particle. Using this "charged point particle dancing in a potential well" to describe the static quantum particle has proven to be unsuccessful in explaining rather "simple properties" like the particle density or its magnetic moment. This has gone so far that the magnetic moment of the proton, $2.79 \frac{ e \hbar }{2m_{p}}$ ($m_{p}$ being the proton mass) is called "anomalous", because it is shifted 2.79 times from the expected value, i.e. $\frac{e \hbar }{2m_{p}}$. It is now known that the nucleon is not a point charge \cite{Hofstadter} but its description is still attempted via a wave equation \cite{forest}. With all the respect that the wave equation certainly has, this is but half of the explanation of the quantum particle's existence and behavior. Therefore, it is a fundamental error to use the wave equation to describe all the quantum particle's properties, particularly if the quantum particle is not moving. This error has produced enormous controversy because the moving charged point particle should not radiate when it is in the "orbit" \cite {eisberg130} and should not produce an induced alternate current in a fine wire approached to the nucleon or electron \cite{chemist electron}. These points highlight the need for a better model, one which describes the non-moving particle as well as the wave behavior described by the wave equation.\\
A special quantum particle, the nucleon, will not only have a constant magnetic moment, but will also have a toroidal dipole moment or anapole moment, as well as charge, volume, mass and density. The wave equation  is simply unable to explain the origins of these properties.\\
Due to the particle-wave duality \cite{eisberg88}, quantum mechanics has claimed that it is impossible to visualize the particle, which is very understandable if the particle is a four dimensional entity which prints a wave in our universe as it travels trough it. As it has been reported \cite{yepez} a new model that explains quantum properties of the non-moving particle and includes its waving behavior, has expressly stated that the quantum entity must be a hypertorus (a 4-dimensional torus), the time intersection of which with a lower dimensional world (a 3-dimensional space) produces said quantum properties. One of the intersections of a hypertorus would be a 3-dimensional torus. As has been well reported \cite{forest}, femtometer toroidal structures have been detected for the deuteron. In this paper, it is shown why these femtometer toroidal structures occur. What is the magnetic moment of the deuteron? Why do different structures produce the same magnetic moment? What structure will produce an anapole moment instead of a magnetic moment? How does the energy of the particle depend on its geometry? And how do these different structures produce the same mass density?

\section{Geometry and Intersection}
\begin{figure}
\begin{center}
\includegraphics[width=3in]{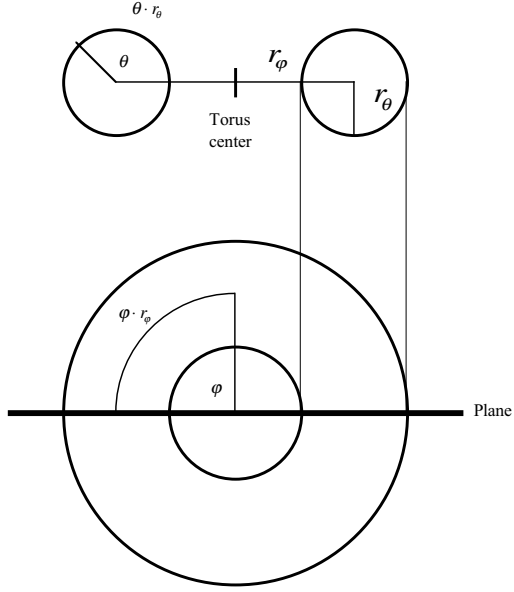}
\caption{Cross section of a 3D Torus intersecting a plane (bottom) and its orthogonal projection (above),  showing its geometric parameters r$_{\theta}$ and r$_{\varphi}$. The arcs $\varphi r_{\varphi}$ and $\theta r_{\theta}$ are also shown. }
\end{center}
\end{figure}
As it has been explained \cite{yepez}, the key hypothesis of this work is that the quantum particle's properties are the consequence of the intersection of a higher dimensional torus (4D) in a lower dimensional world (3D), where the measurement is performed. The technique for grasping how a 4D object would be is to envision the intersection of its 3D version with a 2D space (a plane) and then revolve the resultant through a given edge \cite{beyond the third}. For example, the intersection of a 4D sphere through a 3D space will begin as a tiny sphere that increases in size up to a certain limit (the radius of the 4D sphere) and then reduces its size until it disappears. The equivalent operation between a 3D sphere and a 2D plane will be a tiny circle that increases up to a maximum size (the radius of the sphere) and then reduces its size until it disappears. Therefore, it is easy to visualize that the revolution of this circle will give the equivalent sphere of the previous "impossible to imagine" operation. This is true for the sphere which is highly symmetrical, as any given rotation edge will give the same result. In the case of a 4D torus intersecting the 3D space, however, there is a wide number of possible intersections with the 3D space. One is only concerned with the possible intersections that gives the femtometer toroidal structures found for the deuteron in reference \cite{forest}.\\
In Fig. 1 the cross section of the intersection of a torus with a plane and its orthogonal projection are depicted. Also shown are the geometric parameters that define a torus and the arcs that are going to be used in this paper. In Fig. 2 the result of the operation described in the previous paragraph, for a torus that is intersecting the plane through its $\theta$ circles can be appreciated, just as shown in Fig. 1. It is clearly observed that some of the structures reported in Fig. 6 of ref.\cite{forest} are reproduced. 
\begin{figure}
\begin{center}
\includegraphics[width=4.5in]{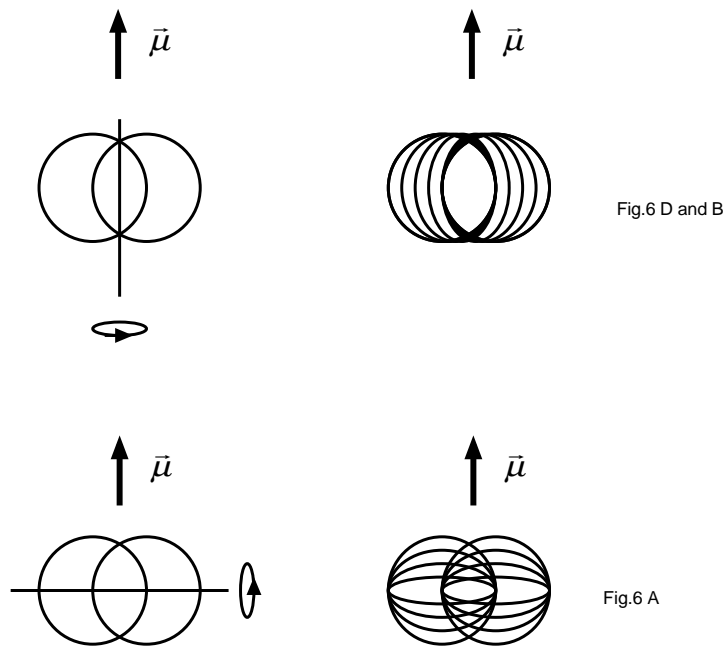}
\caption{Intersection of a 3D torus with the plane (the paper sheet) and the resultant revolution through a polar edge (top) and an equatorial edge (bottom). The revolution of these two projections is on the right side of the figure. The polar revolution will produce a spindle torus which is related to the surfaces B and D, whereas the equatorial revolution is related to the surface A for the deuteron in Fig. 6 reference \cite{forest}.}
\end{center}
\end{figure}

\section{Magnetic Moment and Charge}
As has been explained \cite{yepez}, the origin of the magnetic dipole moment of the quantum particle is the consequence of the time consuming intersection of the particle's toroidal dipole moment (anapole moment) with the plane (2D space or 2D vacuum). The toroidal dipole moment is \cite{dubovic},
\begin{equation}
T=\frac{1}{4\pi c}\cdot IV
\end{equation}
where c is the speed of light and \textit{I} is the current in the toroid' coils, which is,
\begin{equation}
I=\eta i_{\theta}
\end{equation}
where $\eta$ is the number of spires in the torus and  V is the torus's volume,
\begin{equation}
V=2\pi^2r_{\theta}^2r_{\varphi}
\end{equation}
where $r_{\theta}$ is the radius of the cylindrical body of the torus and $r_{\varphi}$ is the distance from the center of the torus to the center of said cylindrical body (see Fig. 1).
Going into more detail of the description given in the first paper \cite{yepez}, an intersection time of such a cylindrical body through a plane will produce circle sectors (see Fig. 1 and 3). The area of a circle sector is,
\begin{equation}
S_{\circ}= \frac{1}{2} \cdot \theta r_{\theta}^{2}
\end{equation}
This circle sector's charge will be the current that pass through the $\theta$ perimeter multiplied by the time to make the arc $\theta r_{\theta}$ at the speed \textit{c}, i.e.,
\begin{equation}
\frac{\theta r_{\theta}}{c}\cdot i_{\theta}
\end{equation}
Given that the surface of the torus is,
\begin{equation}
S=4\pi^2r_{\theta}r_{\varphi}
\end{equation}
and assuming that the charge is uniformly distributed on that surface, the following surface to charge relation can be established,
\begin{equation}
\frac{4\pi^{2}r_{\theta}r_{\varphi}}{e}=\frac{\frac{1}{2} \cdot \theta r_{\theta}^{2}}{\frac{\theta r_{\theta}}{c}\cdot i_{\theta}}
\end{equation}
thus,
\begin{equation}
i_{\theta}=\frac{ec}{8\pi^{2} r_{\varphi}}
\end{equation}
The number of 3D torus spires can be obtained by multiplying this current by the time needed to obtain the charge, \textit{e}, then converting this time to a distance and dividing this result by the $\theta$ perimeter, i.e.,
\begin{equation}
i_{\theta}=\frac{ec}{8\pi^{2} r_{\varphi}}=\frac{e}{t}=>t=\frac{8\pi^{2}r_{\varphi}}{c}=>d=8\pi^{2} r_{\varphi}=>\eta=\frac{8\pi^{2} r_{\varphi}}{2 \pi r_{\theta}}= \frac{4\pi r_{\varphi}}{r_{\theta}}
\end{equation}
Finally, the toroidal moment (anapole moment) is,
\begin{equation}
T= \frac{e r_{\theta}r_{\varphi}}{4}
\end{equation}
Since the intersection time was defined in (5) as $\frac{\theta r_{\theta}}{c}$, the magnetic moment will be,
\begin {equation}
\mu=\frac{\frac{e r_{\theta}r_{\varphi}}{4}}{\frac{\theta r_{\theta}}{c}}=\frac{ec r_{\varphi}}{4\theta}
\end{equation}
The charge of the deuteron nucleus will be the current in (8) multiplied by the time $t_{e}$ needed for all the spires sectors to contribute to the production of the total charge \textit{e},
\begin{equation}
e= i_{\theta}\cdot n \cdot t_{e} =\frac{ec}{8\pi^{2} r_{\varphi}} \cdot  \frac{4\pi r_{\varphi}}{r_{\theta}} \cdot t_{e} => t_{e}=\frac{2\pi r_{\theta}}{c}
\end{equation}

\section{Energy and Mass density}
In the process of intersecting the plane, the $\theta$ perimeter moves a distance $\theta r_{\theta}$, which will correspond to a movement in the $\varphi$ perimeter equal to $\varphi r_{\varphi}$ (see Fig. 1). In the case of an electron in an electric field forming a hydrogen atom, the speed at which its $\varphi$ perimeter is moving is \textit{c} and the speed at which its $\theta$ perimeter is moving is $\frac{e^2}{2h\epsilon_{0}}$, therefore,
\begin{figure}
\begin{center}
\includegraphics[width=3.5in]{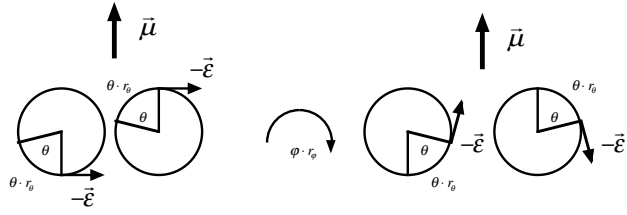}
\caption{Movement of the arc $\theta r_{\theta}$ when the arc $\varphi r_{\varphi}$ occurs.}
\end{center}
\end{figure}
\begin{equation}
\frac{\theta r_{\theta}}{\varphi r_{\varphi}}=\frac{\frac{e^2}{2h\epsilon_{0}}}{c}
\end{equation}
This equation can be rearranged to,
\begin {equation}
\frac{\theta r_{\theta}}{\varphi r_{\varphi}}=\frac{\frac{e^2 4\pi r_{\varphi}}{4\pi \epsilon_{0}r_{\varphi} 2h}}{c}=\frac{2\pi r_{\varphi}E_{\varphi}}{hc}
\end{equation}
where $E_{\varphi}$ is the energy of the toroidal particle with a given $r_{\varphi}$ radius, i.e., the toroidal particle is occupying the whole circle perimeter defined by $r_{\varphi}$ (see Fig. 1). Finally, the energy of the particle is,
\begin {equation}
E_{\varphi}= \frac{\theta r_{\theta}}{\varphi r_{\varphi}^2} \cdot \hbar c
\end{equation}
Even though the energy equation deduced would be related to a particle under an electric field, it also reproduces the particle mass energy, depending on the torus geometric parameters, as if it were a state function.\\
\begin{figure}
\begin{center}
\includegraphics[width=4in]{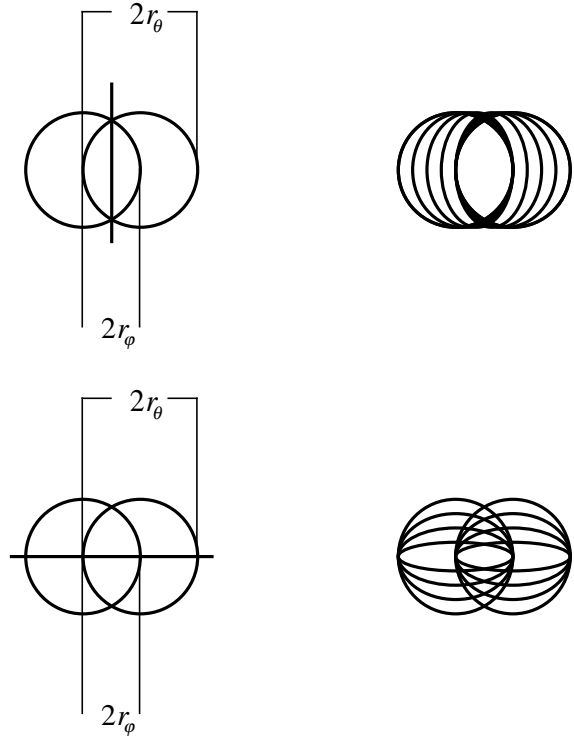}
\caption{Measurements of $r_{\theta}$ and $r_{\varphi}$ from the torus intersection with the plane}
\end{center}
\end{figure}
Once the equation for the energy is obtained, its mass can be easily calculated simply by dividing by $c^{2}$,
\begin {equation}
m_{\varphi}= \frac{\theta r_{\theta}}{\varphi r_{\varphi}^2} \cdot \frac{\hbar}{c}
\end{equation}
It follows that the density of the particle is,
\begin {equation}
\rho= \frac{\frac{\theta r_{\theta}}{\varphi r_{\varphi}^2} \cdot \frac{\hbar}{c}}{2\pi^2r_{\theta}^2 r_{\varphi}}=\frac{\theta}{2 \pi^2 \varphi r_{\theta} r_{\varphi}^3}\cdot \frac{\hbar}{c}
\end{equation}
In Table 1, the measurements done on the toroidal structures reported for the deuteron in ref. \cite{forest}, using the procedure depicted in Fig. 4 are reported.
\begin{figure}
\begin{center}
\includegraphics[width=4in]{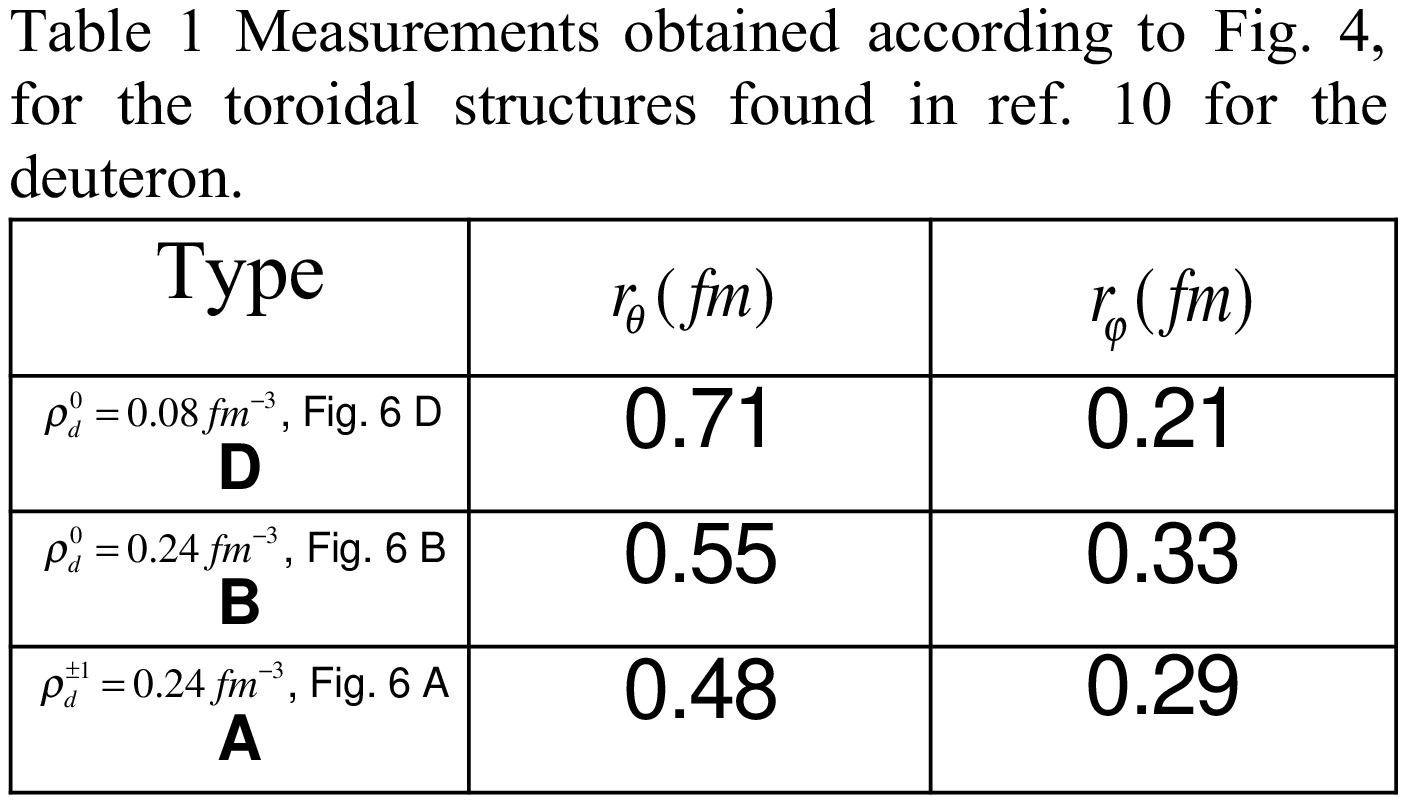}
\end{center}
\end{figure}
In Table 2, the results of using equations (11), (15) and (17) provide the magnetic moment, the energy and the density of the deuteron respectively.\\ 
\begin{figure}
\begin{center}
\includegraphics[width=4in]{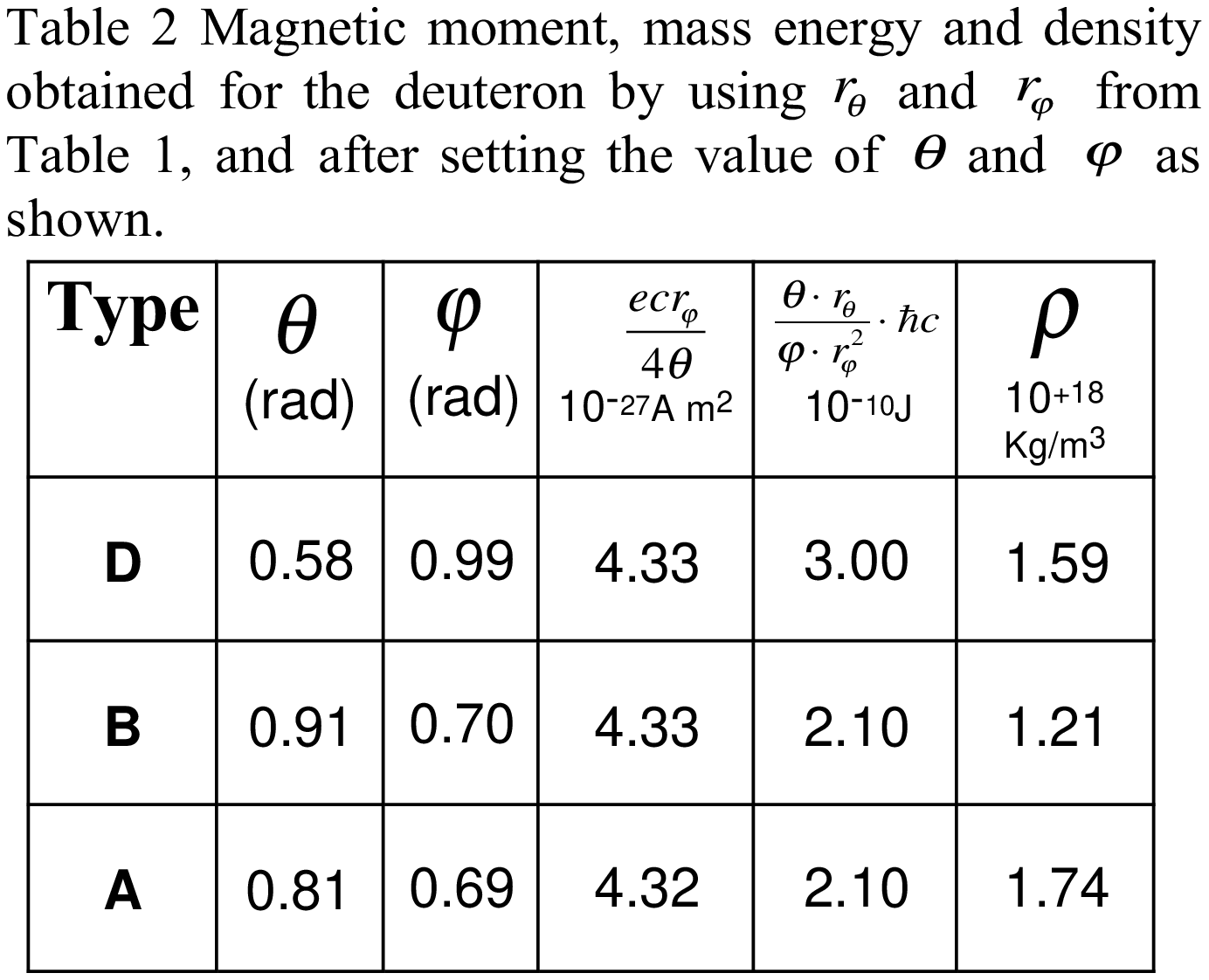}
\end{center}
\end{figure}
\begin{figure}
\begin{center}
\includegraphics[width=4in]{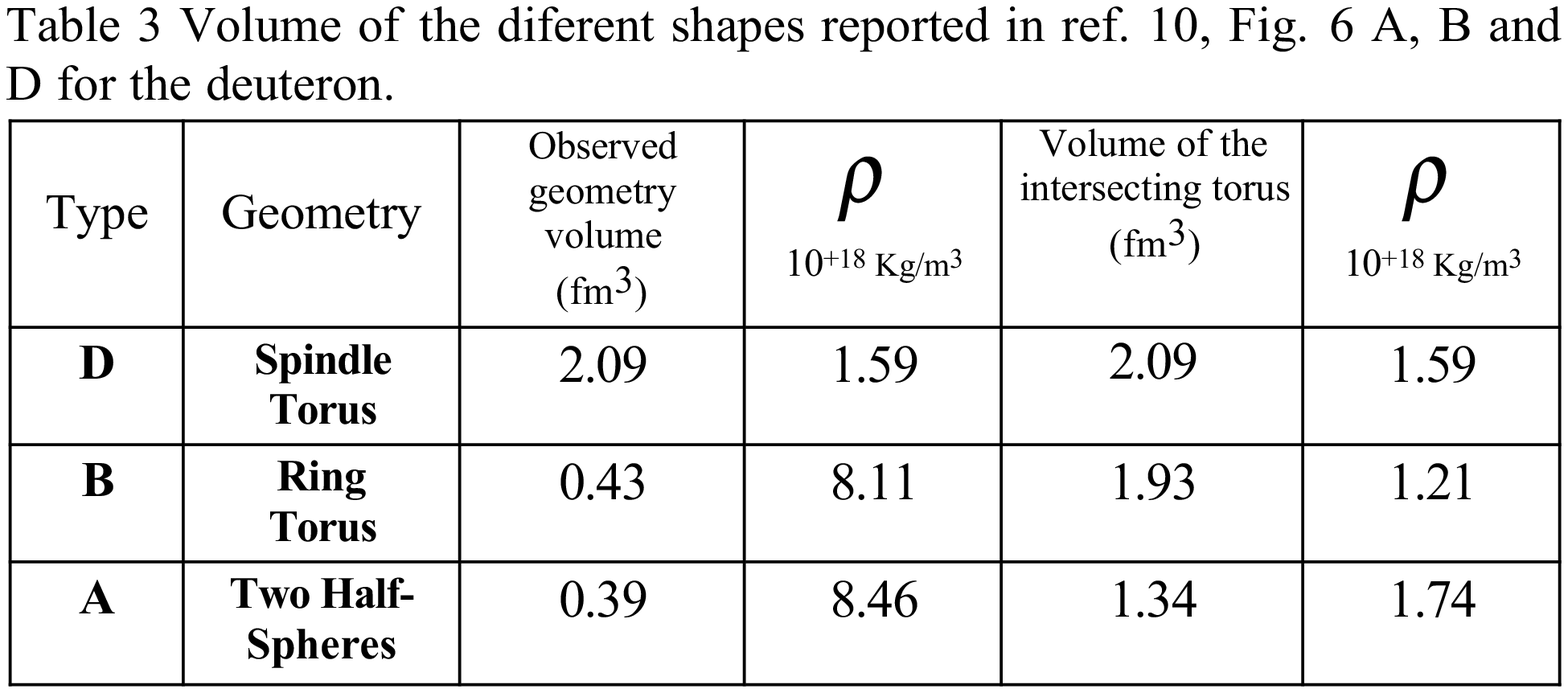}
\end{center}
\end{figure}

\section{Discussion}
The strong resemblance found between the revolution of the intersection left in the plane shown in Fig. 2 and the reported femtometer toroidal structures A, B and D \cite{forest} is more than a mere coincidence. 
This intersection is the only way that a time-dependent anapole moment (a magnetic moment) can be left by the torus in the plane, just as it has been described \cite{yepez}. In addition it is required that, regardless of their shapes, all these structures must produce the same magnetic moment.
The spindle torus reported as Fig. 6 D, as well as the two disconnected semi-spheres reported as Fig. 6 A, both in ref. \cite{forest}, are  clearly reproduced (top and bottom respectively). The central core observed in both structures is due to the intersection of two "separated" sectors of the intersecting torus as shown and therefore, no tensor force \cite{forest}  is needed to explain it.
As is seen in Fig. 3, the same mechanism that produces the magnetic moment of the particle also produces the electric field vectors in the plane and the waving behavior of the particle, when the particle travels through it, as it has been explained in ref. \cite{yepez}. Therefore, if one assumes that the particle is a wave, as it is the current belief, the Dirac and/or the Shr\"odinger equation will describe the wave but will not explain how a static particle's magnetic moment is produced. In this regard, the current belief not only falls short but is also incapable of describing such behavior. The present model on the contrary offers the possibility of obtaining the static particle's properties while its waving behavior remains secondary.\\
The deduction of the magnetic moment is relatively simple given the guidance reported \cite{yepez}, where a circular sector is clearly proposed but, at that time, was not elaborated further. The charge of the whole $\theta$ circle and another intersection time was used instead.
Excellent agreement between equations (11) and the experimental value for the magnetic moment for the deuteron shapes reported as Fig. A, B and D of ref. \cite{forest} were found, not only in the order of magnitude but also in the absolute value. Using all the values obtained in Table 2, it was possible to measure 4.331 $^{+}_{-}$  0.005 10$^{-27}$ A m$^2$, in perfect agreement with the experimental value \cite{deuteron}. Therefore, it appears that the mechanism to produce this magnetic moment has been unveiled. This result, has not been achieved by the current models, which give no indication on how the magnetic moment of the deuteron is produced.\\
The $\theta$ and $\varphi$ values found are arbitrary, but in the case of $\theta$ those values can not be larger than 1 radian (57$^\circ$), as larger values will not produce a wave when the particle travels through the plane as it has  been described \cite{yepez}. Given that all the $\theta$ values were below or equal to 1 radian, they can be used in confidence. The same happens with $\varphi$ values; the fact that they are not bigger than 1 radian makes them reliable. This is true in the sense that the intersecting torus is not doing a complete round in the $\varphi r_{\varphi}$ arc ($2\pi$ radians) to produce the magnetic moment vectors, but rather a fraction of it, which corresponds to a fraction of the arc $\theta r_{\theta}$.\\
The equation for the energy of the free particle (15) was found as a new interpretation of the hyperfine structure constant, expressed as the ratio of two speeds (equation 13). However, the fact that this equation also reproduces the mass energy of the particle, when the geometric parameters of structure D (see Table 1 and 2) were used, is telling us that equation (15) is a function of state, i.e. it does not matter if the particle is under an electric field or free, its energy just depends on the kind of particle and how much space it covers. This was clearly suggested intuitively in \cite{yepez}: the energy of the particle depends on the tension that the particle exerts on the space. This tension is directly proportional to the magnitude of the electric field vectors left in the plane, which happens each  $\frac{\theta r_{\theta}}{c}$, hence this direct proportionality. And it is inversely proportional to the square of the distance $r_{\varphi}$, which is logical as more area means less tension. This is why the lower energy structures A and B have larger $r_{\varphi}$ than structure D (see Table 1).\\
As it can be appreciated in Table 2, just the structure D produced the mass energy of the deuteron \cite{eisberg597}, therefore, this structure could be regarded as the static state of the deuteron as Forest et al. suggested, based on other reasons \cite{forest}. 
The other two structures presented a lower energy but the particle density was kept constant. By using all the density values obtained, it was possible to measure a density of 1.5 $^{+}_{-}$  0.3 10$^{+18}$ Kg/ m$^3$. This value is more precise than the same density value obtained by using the observed geometries reported in ref. \cite{forest} (see Table 3) and keeping the mass of the deuteron constant (3.35 10$^{-27}$ Kg), giving 6 $^{+}_{-}$  4 10$^{+18}$ Kg/ m$^3$. This is because the value of nuclear density given by the structures A and B, in this exercise, is quite unreliable ($\sim$ 8 10$^{+18}$ Kg/ m$^3$). Therefore, the diminution of the energy obtained through equation (15) is needed to keep the mass density constant. 
The Yukawa potential \cite{eisberg722}, was used also, instead of equation (15), but the average density values obtained for the structures A, B and D is 4 $^{+}_{-}$  1 10$^{+18}$ Kg/ m$^3$, which is less precise and quite unreliable. These values were obtained after using $r^{\prime}=\frac{\hbar}{m_{\pi}c}\simeq 1.5 fm$. where $m_{\pi}$ is the meson mass. Only after setting the $r^{\prime}$ in the Yukawa potential to the $r_{\varphi}$ for each structure, the average nuclear density (structures A, B and D) is 2.0 $^{+}_{-}$  0.5 10$^{+18}$ Kg/ m$^3$, which is as good as the density reported above, but happens because the Yukawa potential converges to equation (15) when  $r^{\prime}=r_{\varphi}$.\\
If one uses a sphere as a model, the nuclear mass density of the deuteron would be 2.8 10$^{+17}$ Kg/ m$^3$ (using 1.42 \textit{fm} as the sphere radius, two times D's $r_{\theta}$). This value is very close to the accepted average value of nuclear mass density, 2.3 10$^{+17}$ Kg/ m$^3$ \cite{nuclear mass density}. It is obtained, evidently, because a wrong model is in use. The ratio between the volume of a sphere with a diameter equivalent to the external diameter of the torus and the volume of said torus in the case of the structure D is 6. Hence, the nuclear mass density is lower than that reported above for the deuteron.\\
\begin{figure}
\begin{center}
\includegraphics[width=3in]{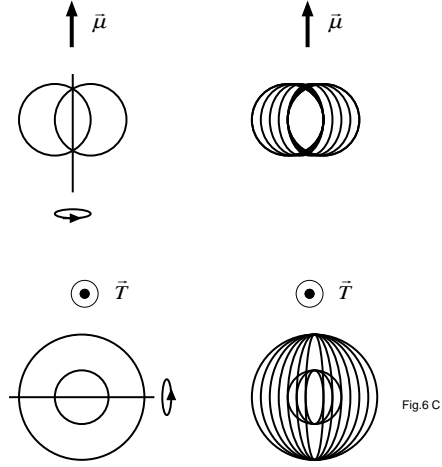}
\caption{Production of the magnetic moment (above) and the anapole moment (bottom). The latter clearly reproduces Fig. 6 C in ref. \cite{forest}.}
\end{center}
\end{figure}
It has been suggested \cite{forest} that the deuteron structure A is the result of a rotating torus. However, if that is so, its density would be unreliable (the density of A would be around 4 10$^{+18}$ Kg/ m$^3$ producing a mean value of 5 $^{+}_{-}$  3 10$^{+18}$ Kg/ m$^3$ ). All this leads to the realization that the intersecting torus is what is important and not its projections in the space. As it is clearly expressed, mass (a form of energy) has lost its absoluteness; it is the mass density that is absolute.\\
According to equation (12), the deuteron yields a charge equal to \textit{e}, after all the spires of the intersecting torus contribute to the charge, i.e. the existence of the particle is not complete until the time $t_{e}$ has passed. Since this time is 7-12 times higher than the time needed to produce the magnetic moment, just partial existence of the particle is needed to produce it. This result is not surprising, because it is produced when the charge \textit{e} is uniformly distributed over all of the intersecting torus surface. However, the charge involved in the production of the magnetic moment was the charge of the circle sector and since the current for both processes is the same, there should be a difference between both times.\\
In other to avoid having two different intersection times, the circle sector must posses all the charge \textit{e}. Therefore, equation (7) becomes,
\begin{equation}
\frac{4\pi^{2}r_{\theta}r_{\varphi}}{\frac{8\pi^2r_{\varphi}e}{\theta r_{\theta}}}=\frac{\frac{1}{2} \cdot \theta r_{\theta}^{2}}{\frac{\theta r_{\theta}}{c}\cdot i_{\theta}}
\end{equation}
thus,
\begin{equation}
i_{\theta}=\frac{ec}{\theta r_{\theta}}
\end{equation}
the number of spires would be,
\begin{equation}
i_{\theta}=\frac{ec}{\theta r_{\theta}}=\frac{e}{t}=>t=\frac{\theta r_{\theta}}{c}=>d=\theta r_{\theta}=>\eta=\frac{\theta r_{\theta}}{2 \pi r_{\theta}}= \frac{\theta}{2\pi}
\end{equation}
rendering equation (10) again. The fact that now there is a charge different from \textit{e} evenly distributed on the surface of the intersecting torus is not a problem since it would be a hyper-charge, i.e. this charge will not be in the plane, and \textit{e} will only occur during the intersection event. The number of spires $\eta$ is now a fraction of a circle, which is actually what is proposed to happens and it is clearly observed on Fig. 3: \textbf{\textit{"a current is intersecting in the plane during a given time, producing the charge of the particle and the electric field vectors of its wave. The same current is toroidal in shape producing the anapole moment of the particle, which is independent of time. When the intersection occurs with a given orientation (which is the same needed to produce the particle's wave), that anapole moment becomes time dependent and, therefore, the particle magnetic moment appears. The electric field vectors tense the space producing the particle mass or energy, the mass energy and volume of the torus change to keep its density constant"}}\\
One may think that all that has been described is a 3D torus, with an intermittent toroidal current. However, such a device will never produce a magnetic moment, will not have a net charge and will not change its mass energy if it is deformed, etc. Ergo, all those properties happen as a consequence of a time consuming interaction of the particle with the space, which is very different and strongly fundamental.\\
As observed, the anapole moment of the particle is the source of its magnetic moment. Since the deuteron exhibits both momenta, its anapole moment will be produced when the torus is intersecting the plane across its equator, as observed in Fig. 4, which clearly reproduces the structure observed as Fig. 6 C in ref. \cite{forest}. Hence, that structure will not produce a magnetic moment, whereas the other three will. It would not matter that the anapole moment is projected out side the plane, as its effect will be felt but no magnetic moment will be measured in this way of intersection. Using this interpretation, structure C's  $r_{\theta}$ is 0.45 \textit{fm} and $r_{\varphi}$ is 0.18 \textit{fm}, which produces an anapole moment of 3 10$^{-47}$ C cm$^{2}$.
This loss of the magnetic moment as a consequence of a change in the way the particle is intersecting the plane is the key to understanding EPR paradox, because this way of intersection produces the loss of separation by class of spin of particles achieved with magnetic filtering, as has been fully discussed \cite{yepez}.\\
The theory for the existence of the anapole moment was developed for atoms, using the shell model. It uses the spin of a s$_{\frac{1}{2}}$ state. If that spin is directed along the z axis, then with increasing distance from the origin in the xy plane, the spin acquires a projection on this plane along the tangent to the circle. The resulting configuration is just a spin helix \cite{khriplovich17}, which will have a handedness and therefore, will explain the parity nonconservation of the nucleus. In the present model handedness is intrinsic to the particle and happens as a consequence of the 360$^{\circ}$ twist made in the $\theta$ axis as has been fully described \cite{yepez}.\\
The vector of anapole moment is \cite{khriplovich138},
\begin {equation}
T=-\pi \int dr r^{2} J(r) 
\end{equation}
J(r) is,
\begin {equation}
J(r)=\frac{e \mu}{2m}\nabla\times(\psi\dag\sigma\psi)
\end{equation}
which describe the contribution of the spin current and $\psi(r)$ is,
\begin {equation}
\psi(r)=(1-i \frac{G}{\sqrt{2}}g\rho_{0}\sigma r) \psi_{0}(r)
\end{equation}
finally, the anapole moment for the nucleus is,
\begin {equation}
T=\frac{G}{\sqrt{2}}\frac{2\pi e \mu}{m_{p}} g \rho_{0} < r^{2} > \frac{KI}{I(I+1)},   K=(I+\frac{1}{2})(-1)^{\frac{I+1}{2-l}}
\end{equation}
Here, G is the Fermi weak interaction constant and \textit{g} is the natural scale of atomic P-odd effects. $< r^{2} > $ is the mean squared radius of the external nucleon. It coincides to good accuracy with the squared charge radius of the nucleons, $r_{1}^{2}$=$\frac{3}{5}r_{0}^{2}$=$\frac{3}{5}\widetilde{r}_{0}^{2}A^{\frac{2}{3}}$, $\widetilde{r}_{0}= 1.2 10^{-13}$ cm. Setting $\rho_{0}=(\frac{4\pi}{3}\widetilde{r}_{0}^{3})^{-1}$, we finally obtain:
\begin {equation}
T=\frac{9 G g}{10 \sqrt{2}} \frac{e \mu}{m_{p}\widetilde{r}_{0}} A^{\frac{2}{3}} \frac{KI}{I(I+1)}
\end{equation}
Using equation (22) the anapole moment for the deuteron is, $\sim$ 3 10$^{-48}$ C cm$^{2}$, in line with the rough estimate gave by Zel'dovich, i.e. $10^{-26}\mu \sim 10^{-49}$ \cite{zeldovich}. However, this value has to be corrected because it was done for a sphere of radius 1.2 \textit{fm}. Multiplying by the volume of that sphere and dividing between the volume of the torus of structure C, the anapole moment predicted by equation (22) for the deuteron gives $\sim$ 3 10$^{-47}$ C cm$^{2}$, which is in excellent agreement with the anapole value obtained in this paper. After obtaining equation (22) the author acknowledged that the anapole corresponds to a magnetic field configuration induced by a toroidal winding. It is clear that the anapole moment must be proportional to the magnetic flux, i.e., to the cross-section area of the torus! \cite{khriplovich138}. This is in direct agreement with the model presented here. The existence of the anapole moment implies a toroidal winding current, therefore there should be a timing mechanism to stop such current at some time in order to produce a constant charge, the intersection time comes to fulfill this requirement.\\
Finally, this model implies that a homogeneous torus, i.e. the neutron and proton spires, are interwoven in the deuteron torus, i.e., no separation of particles in the nucleus is needed.

\section{Conclusions}
Assuming that the quantum particle is not a plane wave, a better description of a quantum particle is provided, where the particle wave behavior is also included. This is consistent with the particle-wave duality, where the wave interpretation has prevailed over the particle interpretation. Evidence that supports the more fundamental idea that the quantum particle properties are the consequence of the intersection of a higher dimensional object in a lower dimensional world is provided. Mass is not an absolute quantity and what could be related to the absoluteness of mass is better described by the particle mass density, which was found to be constant regardless of the changes in the mass energy of the particle. Evidence that the way of intersection with the space will produce either an anapole moment or a magnetic moment, but not both, is provided, supporting the explanation of EPR paradox discussed in ref. \cite{yepez}, where the loss of the magnetic moment on a given orientation of the particle is the key to explaining it. More research like that published by Forest et al. \cite{forest} for the deuteron is needed to corroborate these findings with other quantum particles.

\section{Acknowledgment}
The author is thankful to Mr. Aaron Rowe and Mrs. Wanda Aylward for useful corrections of the manuscript.

 \end{document}